\documentclass[12pt]{iopart}

\expandafter\let\csname equation*\endcsname\relax
\expandafter\let\csname endequation*\endcsname\relax
\usepackage{amsmath}

\usepackage{iopams}
\usepackage{graphicx}
\usepackage{amssymb}
\usepackage{dsfont}
\usepackage{epstopdf}
\usepackage{color}
\usepackage{cancel}

\usepackage{mathtools}
\usepackage{hyperref}

\definecolor{red}{rgb}{1,0,0}
\definecolor{blue}{rgb}{0,0,1}
\definecolor{black}{rgb}{0,0,0}

\newcommand{\p}{\partial}
\newcommand{\eq}[1]{\begin{align}#1\end{align}}
\newcommand{\eqs}[1]{\begin{align*}#1\end{align*}}
\newcommand{\ffrac}[2]{\mbox{$\frac{#1}{#2}$}}
\newcommand{\half}{\mbox{$\frac{1}{2}$}}
\newcommand{\OO}{\mathcal{O}}
\newcommand{\GG}{\mathcal{G}}

\newcommand{\PP}{\mathbb{P}}
\newcommand{\ZZ}{\mathbb{Z}}

\newcommand{\FF}{\mathbb{F}}
\newcommand{\TT}{\mathcal{T}}

\newcommand{\lbar}{\ell_0}

\newcommand{\Mbar}{{\overline M}}
\newcommand{\Obar}{{\overline O}}

\newcommand{\delb}{{\hat \delta}}

\makeatletter
\g@addto@macro\@floatboxreset\centering
\makeatother

\begin{document}
\title{Emergence of order in random languages}
\author{Eric De \!Giuli} 
\address{Institut de Physique Th\'eorique Philippe Meyer, \'Ecole Normale Sup\'erieure, \\ PSL University, Sorbonne Universit\'es, CNRS, 75005 Paris, France}
\ead{degiuli@lpt.ens.fr}

\begin{abstract}
We consider languages generated by weighted context-free grammars. It is shown that the behavior of large texts is controlled by saddle-point equations for an appropriate generating function. We then consider ensembles of grammars, in particular the Random Language Model of \cite{DeGiuli19}. This model is solved in the replica-symmetric {\it ansatz}, which is valid in the high-temperature, disordered phase. It is shown that in the phase in which languages carry information, the replica symmetry must be broken.
\end{abstract}

\noindent{\it Note: The body is this work is as published in J. Phys. A: Math. Theor. 52 (2019) 504001. A Corrigendum has been added as Appendix B} \\

\begin{figure}[t!]
\includegraphics[width=0.7\textwidth]{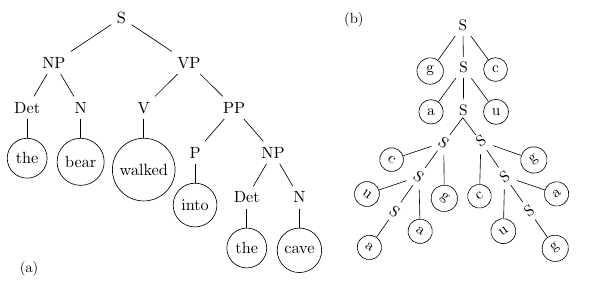}
\caption{ Illustrative derivation trees for (a) simple English sentence, and (b) RNA secondary structure (after \cite{Searls02}). The latter is a derivation of the sequence `gacuaagcugaguc' and shows its folded structure. Terminal symbols are encircled. 
}\label{fig1}
\end{figure}

Many complex systems have a generative, or linguistic, aspect. For example, protein structure is written in sequences of amino acids, a language of 20 different symbols. { A large body of previous work has investigated the social aspect of linguistic systems, namely that different agents must find consensus regarding the meaning of symbols \cite{Loreto07,Loreto11,Burridge17}. A complementary but necessary aspect of any linguistic system concerns the hidden structure within the sequences themselves, independent of communication. The most basic structural property is syntax: the rules that govern how symbols can be combined to create richer structures and thus carry information. In computer science and linguistics, generative grammar has proved to be a valuable formalism to describe syntax, in a generalized sense} \cite{Carnie13,Chomsky02,Hopcroft07}. A generative grammar consists of an alphabet of hidden symbols, an alphabet of observable symbols, and a set of rules, which allow certain combinations of symbols to be replaced by others. From an initial start symbol S, one progressively applies the rules until only observable symbols remain; any sentence produced this way is said to be `grammatical,' and the set of all such sentences is called the language of the grammar. The sequence of rule applications is called a derivation. The Chomsky hierarchy distinguishes grammars based on the complexity of the grammatical rules. In this work, we restrict our attention to context-free grammars (CFGs), for which derivations are trees (Figure \ref{fig1}). 

\begin{figure}[t!]
\includegraphics[width=0.5\textwidth]{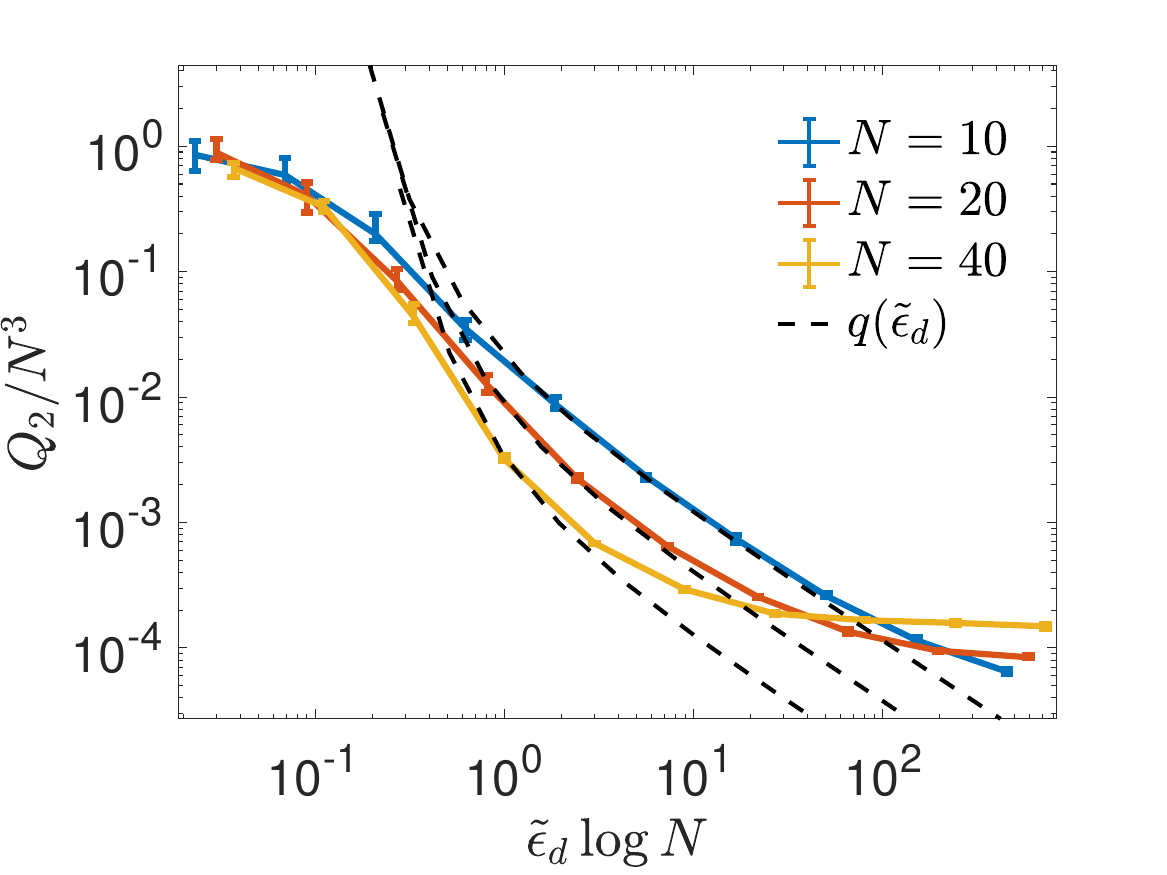}
\caption{Order parameter $Q_2$ on logarithmic axes. Solid lines show numerical data from random grammars with $N$ as indicated and $\ell \approx 10^5$. The plateau at large $\tilde \epsilon_d$ is a finite-$\ell$ effect; empirically it scales as $Q_2^\infty \sim N^4/\ell$. The function $q(\tilde\epsilon_d) = (e^{1/(2\tilde \epsilon_d)}-1) (N^2-1)/N^4$ is the theoretical prediction, Eq.\ref{Q2}. }\label{figQ2}
\end{figure}

There are many theoretical results on the capabilities of CFGs \cite{Hopcroft07}.  However, little is known about the statistical properties of {\it large, typical} grammars. { Recently, there has been increasing interest in approaching the properties of syntax from the point of view of statistical \cite{Lin17,DeGiuli19} and quantum \cite{Piattelli-Palmarini15,Gallego17,Pestun17} physics.} In \cite{DeGiuli19}, a simple model of random languages was proposed, using weighted context-free grammars. (We refer the reader to this work for all necessary details about generative grammars). With numerical simulations it was shown that as a `temperature' $\epsilon_d$ is lowered, the model has a transition from a `disordered' phase in which the grammar just produces noise, to an `ordered' phase in which the grammar produces sequences with a nontrivial structure. This transition is characterized by emergence of order in the hidden structure, as measured by an order parameter $Q_2$, defined below. As the temperature is lowered through $\epsilon_*$, $Q_2$ rapidly increases. With a simple scaling argument, the location of this transition was understood as the place where energetic fluctuations begin to be comparable to entropy. However, detailed information on the transition was not obtained in \cite{DeGiuli19}. In this work we use the replica method, as well as diagrammatic methods, to study random languages. We will solve the Random Language Model of \cite{DeGiuli19} in the replica-symmetric phase, and show that it quantitatively captures the behavior of $Q_2$, as shown in Figure \ref{figQ2}. We will see that the replica symmetry must be broken in the ordered phase.

\section{Partition functions}

A CFG in Chomsky Normal form is defined by two types of rules: $a\to bc$, where one hidden symbol becomes two hidden symbols, and $a \to B$, where a hidden symbol becomes an observable symbol. In a weighted CFG (WCFG), we assign weights $M_{abc}$ and $O_{aB}$, respectively, to these rules. With these weights we build a weight for an entire derivation tree, $\TT$, as follows. We write $\sigma$ for the hidden variables, and $o$ for the observables. These are indexed by their positions on the tree. We write $\Omega_\TT$ for the set of internal factors, i.e. factors of the form $a \to b c$, and $\p \Omega_\TT$ for the boundary factors, i.e. those associated to $a \to B$ rules. The number of boundary factors is written $\ell_\TT$, which is also the number of leaves. Since derivations are trees, the number of internal factors is $\ell_\TT-1$. { Write $\GG$ for the objects $M$ and $O$, which specify the grammar. A table of commonly appearing symbols is shown in Table \ref{tab}. }

\begin{table}
  \centering
\begin{tabular}{l | p{6 cm} | p{6 cm} }
Symbol              & Definition & Linguistic Interpretation  \\
\hline
$a,b,\ldots$ & \parbox[t]{6cm}{alphabet of hidden symbols\\(non-terminal symbols)}  & \parbox[t]{6cm}{abstract categories: noun, \\verb, noun phrase, \ldots}  \\
$S$ & start symbol & `sentence' \\
$A,B,\ldots$ & alphabet of observable symbols & words: `the', `bear', `walked', $\ldots$\\
& (terminal symbols) &  \\
$M_{abc}$ & weight for rule $a \to bc$ & grammatical weight \\
$O_{aB}$ & weight for rule $a \to B$ & part-of-speech weight \\
$\sigma_i$ & hidden symbol on site $i$ & \\
$o_i$ & observable symbol on site $i$ & \\
$\pi_{abc}$ & \# of occurrences of rule $a\to bc$ & \\
$\rho_{aB}$ & \# of occurrences of rule $a\to B$ & \\
$N$ & \# of hidden symbols & \# of abstract categories \\
$T$ & \# of observable symbols & \# of words in lexicon \\
$m$ & \# of sentences & \\ 
$\ell$ & \# number of leaves & total number of words \\ 
$\ell_0$ & $\ell-m$ & \\ 
\hline
$L_a$ & downward msg of left-child& \\
$L^\dagger_a$ & upward msg of left-child& \\
$R_a$ & downward msg of right-child& \\
$R^\dagger_a$ & upward msg of right-child & \\
$H_a$ & $L_a^\dagger + R_a^\dagger$ & head message \\
$\eta,\zeta,\xi$ & Lagrange multipliers to control & \\
& \# of branches, roots, leaves & \\
\hline
$\epsilon_d$ & deep temperature & inverse variance of $M$ \\
$\epsilon_s$ & surface temperature & inverse variance of $O$ \\
$\tilde \epsilon_d$ & $\epsilon_d/N^3$ & \\
$\tilde \epsilon_s$ & $\epsilon_s/(NT)$ & \\
\end{tabular}
\caption{Table of symbols. `msg' = message. First block: basic symbols; second block: variables in diagrammatic representation; third block: variables in grammar ensemble.} \label{tab}
\end{table}

Consider a derivation tree, such as that depicted in Figure \ref{fig1}a. In a weighted context-free grammar, a tree  $\TT$ with hidden variables $\sigma_i$ and observables $o_t$ has a weight
\eq{ \label{W1}
W( \{\sigma_i, o_t \} | \TT, \GG) = \prod_{\alpha \in \Omega_\TT} M_{\sigma_{\alpha_1} \sigma_{\alpha_2} \sigma_{\alpha_3} } \prod_{\alpha \in \p\Omega_\TT} O_{\sigma_{\alpha_1} o_{\alpha_2} },
}
where each $\alpha=(\alpha_1,\alpha_2,\alpha_3)$ is a factor in the order $\sigma_{\alpha_1} \to \sigma_{\alpha_2} \sigma_{\alpha_3}$. This defines a conditional probability measure on configurations
\eq{ \label{Pconfig}
\PP( \{\sigma_i, o_t \} | \TT, \GG ) = \frac{W( \{\sigma_i, o_t \} | \TT, \GG) }{Z(\TT, \GG)}
}
where
\eq{ \label{Z1}
Z(\TT,\GG) = \sum_{\{ \sigma_i, o_t \} } W( \{\sigma_i, o_t \} | \TT, \GG)
}
{ Eq.\eqref{Pconfig} specifies the probability of a derivation, given the tree. However, the topology of the tree is itself dynamical. We will consider that trees are chosen as a Bernoulli process: beginning from the root, each hidden variable either becomes an observable, with probability $p$, or branches into two hidden variables, with probability $1-p$. This implies that $\PP(\TT | \GG)  = W_{tree}(\TT)/Z_{tree}$ with $W_{tree}(\TT) = p^{|\p \Omega_\TT|} (1-p)^{|\Omega_\TT|}$. The `emission probability' $p$ controls the size of trees \footnote{$Z_{tree} = 2p/(1+|2p-1|) = 1$ for $p>1/2$.}. }
The tree-averaged partition function is then
\eq{
\ZZ(\GG; \ell) = \sum_{\TT, \ell(\TT)=\ell} \PP(\TT| \GG) Z(\TT,\GG)
}
as a function of the number of leaves $\ell$, and the partition function for $m$ sentences of total length $\ell$ is
\eq{
\ZZ(\GG; m, \ell) = \sum_{\{\ell_i\}, \sum_{i=1}^m \ell_i = \ell} \prod_i \ZZ(\GG; \ell_i) 
}
We have $\sum_{i} |\p \Omega_{\TT_i}| = \sum \ell_i = \ell$, and $\sum_i |\Omega_{\TT_i}| = \sum (2\ell_i-1) = 2\ell - m$, so that $\ZZ_{tree} \equiv \prod_i \PP(\TT_i|\GG)=p^{\ell} (1-p)^{2\ell-m} Z_{tree}^{-m}$ just gives a trivial factor. For now we suppress dependence on $m$ and $\ell$. Note that $\ZZ(\GG; m,\ell)$ is the weight for the grand canonical partition function $\sum_{m,\ell \geq 0} \ZZ(\GG; m,\ell)$; we will, however, work at fixed $m$ and $\ell$. 

\section{Energy, Entropy, and Order Parameters}

It is convenient to add some auxiliary parameters in order to extract additional observables. First, we note that \eqref{W1} can be written as
\eq{ \label{W2}
W( \{\sigma_i, o_t \} | \TT, \GG) = \prod_{a,b,c} M_{abc}^{\pi_{abc}} \; \prod_{a,B} O_{aB}^{\rho_{aB}}
}
where $\pi_{abc}(\sigma)$ is the usage frequency of rule $a\to bc$ and $\rho_{aB}(\sigma,o)$ is the usage frequency of $a\to B$. Adding external fields $h_{abc}$ and $k_{aB}$ let us define the energy of a configuration as
\eq{ \label{E}
E = -\sum_{a,b,c} \pi_{abc} \left[ h_{abc} + \log M_{abc} \right] - \sum_{a,B} \rho_{aB} \left[ k_{aB} + \log O_{aB} \right]
}
Then we can generalize \eqref{Z1} to 
\eq{
Z_{h,k}(\TT,\GG) = \sum_{\{ \sigma_i, o_t \} } e^{-\beta E}
}
where we added a bias $\beta$. The original ensemble is recovered for $\beta=1$ and $h=k=0$. We see that the average energy is
\eq{
\langle E \rangle = -\frac{\p \log Z}{\p \beta}
}
and it is natural to define the entropy of the grammar as $S = \beta \langle E \rangle + \log Z$.

In \cite{DeGiuli19} it was argued that a natural order parameter for WCFGs is one that measures the extent to which rules are applied uniformly: if all rules $a \to bc$ and $a \to B$ have the same weight, the grammar carries no information, and sentences will be indistinguishable from noise. To measure order in the deep grammar, define first
\eq{ \label{Q1}
Q_{abc}(\GG) = \langle \delta_{\sigma_{\alpha_1},a} \big( N^2 \delta_{\sigma_{\alpha_2},b}\delta_{\sigma_{\alpha_3},c} -1 \big) \rangle = \frac{N^2}{\ell_0} \langle \pi_{abc} \rangle - \frac{1}{\ell_0}\sum_{b',c'} \langle \pi_{ab'c'} \rangle, 
}
averaged over all interior vertices $\alpha$, and averaged over derivations. The normalization $\ell_0=\ell-m$ for $\ZZ$. A spin-glass order parameter specific to deep structure is
\eq{ \label{Q21}
Q_2 \equiv \overline{\sum_{a,b,c} Q_{abc}^2} = \frac{N^5(N^2-1)}{\ell_0^2} (q_0-q_1)
}
where
\eq{
q_0 = \overline{\langle \pi_{abc} \rangle^2}, \qquad  q_1 = \overline{\langle\pi_{abc}\rangle\langle\pi_{ab'c'}\rangle},
}
with $b' \neq b, c' \neq c$. Here we used the fact that when $h=0,k=0$, the permutation symmetry is restored upon disorder averaging. We see that
\eq{ \label{pi}
\langle \pi_{abc} \rangle = \frac{1}{\beta} \frac{\p \log Z}{\p h_{abc}}
}
so $q_0$ and $q_1$ can be obtained from derivatives of $\overline{\log Z}$ with respect to the field. 

\section{Diagrammatic formulation}

We expect that universal properties of weighted context-free languages are contained in the behavior of $\overline{\log \ZZ}$ when $\ell$ becomes large, where $\overline{\;\cdot\;}$ is an average over grammars. In order to compute this object, we find it convenient to move to an alternative, particle, representation. In particular, we seek a model whose diagrammatic expansion gives the derivation trees we seek to count, with the appropriate weights. This technique has been widely used in the study of 2D gravity \cite{Di-Francesco95}, and facilitated Kazakov's solution of the Ising model on random surfaces \cite{Kazakov86,Boulatov87}. Later, it was shown in a simpler setting that this technique could be used to easily obtain results for spin models on random graphs \cite{Bachas94,Baillie95}. 

\begin{figure}[t!]
\includegraphics[width=0.8\textwidth]{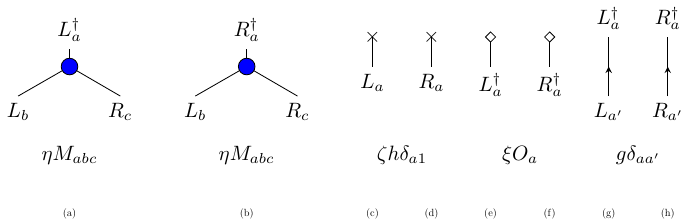}
\caption{Feynman rules for weighted context-free grammars. Elements of Feynman diagram expansion (top row) with weights (middle row). (a-b) $M-$interaction, (c-d) Root source, (e-f) $O$ source, (g-h): Nonzero propagators. Labels $a,a',b,c$ are colour indices, $\in \{1,\ldots,N\}$. All propagators are diagonal in colour.}\label{fig2}
\end{figure}

We begin with a simplified model. Consider the formal integral
\eq{
W = |G|^{-N/2} \int DU \; e^{-\frac{1}{2} \sum_{i,j} U_i G^{-1}_{ij} U_j} \; e^{\xi \sum_i U_i + \eta\sum_{i,j,k} U_i U_j U_k M_{ijk}},
}
where the measure is normalized such that $|G|^{-N/2} \int DU \exp(-\half \sum_{i,j} U_i G^{-1}_{ij} U_j) = 1$. Strictly speaking, the integral is only defined by its perturbative expansion in $\eta$; convergence requires that the real part of $G^{-1}$ be positive-definite. This expansion generates Feynman diagrams with cubic vertices. Each vertex gets a factor $\eta M_{ijk}$, and the expansion with respect to $\xi$ generates sources. By Wick's theorem, each edge gets a factor $G_{ij}$, the propagator. The coefficient in this expansion of $\xi^m \eta^k$ thus counts all such diagrams, possibly disconnected, with $k$ vertices and $m$ sources, times an inverse symmetry factor \cite{Bessis80} \footnote{Such factors appear when diagrams, including all their colour indices, have nontrivial symmetries, like reflections. In the disordered case where $M_{abc}$ depends on all indices, these symmetry factors will not play a role since typical connected graphs will have no symmetries.}.

This is a skeleton of what we need to count derivation trees, but there are several elements missing: first, $W$ includes all graphs with cubic vertices, not only trees. Second, even if we could restrict the sum to trees, there is nothing in $W$ to distinguish leaves from roots, or to distinguish the left and right branches from a given hidden node.

\begin{figure}[t!]
\includegraphics[width=0.3\textwidth]{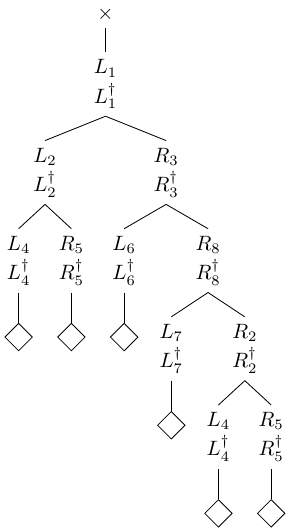}
\caption{Feynman diagram corresponding to derivation tree in Figure 1a. Alphabet of hidden symbols is $\chi_d = (S,NP,VP,Det,N,V,P,PP)$ and alphabet of surface symbols is $\chi_s=(the,bear,walked,into,cave)$. Vertices are represented by $\wedge$ with heads at the tip. The diagram has a weight $2h \zeta \xi^6 \eta^5 g^{11} M_{123}M_{245}^2 M_{368}M_{872} O_{41}^2 O_{52}O_{63}O_{74}O_{55}$.}\label{fig3}
\end{figure}

One solution to these problems is to use matrices as the integration variables, because their diagrammatic expansion can be arranged to give planar diagrams in an appropriate large $N$ limit \cite{tHooft74,Brezin78}. 
 However, for our problem this is overkill. Instead, we will consider a theory of complex scalar fields with colour indices, equivalent to a complex matrix model with matrices of size 1x1. We will have two scalar fields $L_a$ and $R_a$, with colour indices $a= 1 \ldots N$. Consider
\eq{
\FF(\GG) = \int DL \int DR \; e^{-\frac{1}{g} \sum_a \left[ L_a L_a^\dagger + R_a R_a^\dagger \right] } e^{I} 
} 
where ${}^\dagger$ denotes complex conjugate and
\eq{ \label{I}
I = \zeta h (L_1+R_1) + \xi \sum_a O_a (L_a^\dagger + R_a^\dagger) + \eta \sum_{a,b,c} M_{abc} (L_a^\dagger + R_a^\dagger) L_b R_c.
}
with $O_a = \sum_B O_{aB}$. The measure $DL=\prod_a dRe[L_a] dIm[L_a]/(\pi g)$ is normalized such that $\int DL \; e^{-\frac{1}{g} \sum_a L_a L_a^\dagger} = 1$, and similarly for $R$.

The propagator is diagonal in the colour indices $a=1\ldots N$ and for each $a$ is such that $\langle L_a^\dagger L_a \rangle = g$; that is, the Feynman rules are as shown in Figure \ref{fig2}. The diagram corresponding to Figure 1a is shown in Figure \ref{fig3}. We claim that, apart from accidental symmetry factors,
\eq{ \label{W2}
\ZZ(\GG; m, \ell) = m! \oint' \frac{d\zeta}{\zeta^{1+m}} \oint' \frac{d\xi}{\xi^{1+\ell}} \oint' \frac{d\eta}{\eta^{1+\ell-m}} \;\FF(\GG),
} 
where $\oint' = \oint/(2\pi i)$. The proof is as follows. 

The perturbative expansion with respect to $\eta$ generates cubic vertices with distinguished heads and $L$ and $R$ ends; the vertices can be placed on the plane such that that their heads point up. The expansion and contour integral for $\xi$ generate $\ell$ sources of $L^\dagger$ and $R^\dagger$, which are the leaves. The expansion and contour integral for $\zeta$ generate $m$ sources of $L_1$ and $R_1$, which are the roots. An $L_a$ can only be connected to a $L_a^\dagger$, and a $R_a$ can only be connected to a $R_a^\dagger$. We can orient all edges from $L_a \to L_a^\dagger$ and $R_a \to R_a^\dagger$, and similarly for the half-edges in the cubic vertices, flowing from head to L and R branches. Then, from each root, we can define paths by following arrows; any path we take will go through some number of cubic vertices and end in a leaf. Therefore there are $m$ connected components, one for each root. Considered as a graph, we can count the number of edges as follows: each source generates half an edge, and each cubic vertex generates $3/2$ edges. The total number of edges is $\half(m + 3 (\ell-m) + \ell) = 2\ell-m$. The difference between the number of vertices and the number of edges is $(m+\ell+\ell-m)-(2\ell-m) = m$, which is the number of connected components. Therefore the graph is a forest. 



We thus generate a forest of $m$ planar, rooted, trees, with $\ell$ leaves in total. 
The weight of each diagram is
\eq{ \label{diagweight}
h^{m} g^{2\ell-m} \; \prod_{a,b,c} M_{abc}^{\pi_{abc}} \; \prod_{aB} O_{aB}^{\rho_{aB}}
}
where $\pi_{abc}$ is the usage frequency of rule $a\to bc$ and $\rho_{aB}$ is the usage frequency of $a \to B$. This expansion counts diagrams with a degeneracy of $2^m$ since each tree root can be either an $L$ or an $R$. In the expansion, the different connected components are not ordered. We would like to distinguish forests by the order of the trees, so we multiply the result by $m!$. Choosing $g$ and $h$ such that 
\eq{
(2h)^m g^{2\ell-m} = p^\ell (1-p)^{2\ell-m} Z_{tree}^{-m}
}
for all $\ell$ and $m$ we have our result.


The virtue of working with \eqref{W2} is that when $\ell \to \infty, m/\ell = \alpha = $ constant, the leading behavior can be extracted by a saddle-point analysis \cite{Le-Guillou12} \footnote{This can be seen explicitly by considering the rescaling $L = \ell^{1/2} L', R=\ell^{1/2} R', \eta = \ell^{-1/2} \eta', \xi = \ell^{1/2} \xi',\zeta = \ell^{1/2} \zeta'$. }. There is one subtlety. The integration variables are the real and imaginary parts of $L$ and $R$, and the saddle-point equations should be taken with respect to these parts. The solutions to $Re[L_a]$ and $Im[L_a]$, which may be complex, are then added to produce $L_a=Re[L_a]+i Im[L_a]$, and similarly for $R$. By linearity, this is equivalent to taking saddle-point equations with respect to $L_a$ and $L_a^\dagger$, and treating $L_a$ and $L_a^\dagger$ as independent. It is convenient to write $H_a = L_a^\dagger+R_a^\dagger$ for a head. The saddle-point equations are
\eq{
L_a & = g \xi O_a + g \eta \sum_{b,c} M_{abc} L_b R_c \label{SP1}\\
L_a^\dagger & = g h \zeta \delta_{a1} + g \eta \sum_{a',b,c} M_{a'bc} H_a \delta_{ab} R_c \label{SP2} \\
R_a & = g \xi O_a + g \eta \sum_{b,c} M_{abc} L_b R_c \label{SP3} \\
R_a^\dagger & = g h \zeta \delta_{a1} + g \eta \sum_{a',b,c} M_{a'bc} H_a L_b \delta_{ac}  \label{SP4} \\
\ell-m & = \eta \sum_{a,b,c} M_{abc} H_a L_b R_c \label{SP5} \\
\ell & = \xi \sum_{a} O_a H_a \label{SP6} \\
m & = \zeta h (L_1+R_1) \label{SP7}
}
for all $a$.

\eqref{SP1},\eqref{SP2} and their pairs \eqref{SP3},\eqref{SP4} have an interpretation as recursion equations, which are equivalent to the saddle-point limit of Tutte recursion relations or loop equations in related contexts \cite{Di-Francesco06,Eynard16}. They are also related to self-consistent equations derived for spin glasses on trees \footnote{For example, $L_a$ is analagous to what is called $\phi(\{\sigma^a\})$ in \cite{Bachas94} and $g_n(\{\sigma_a\})$ in \cite{Parisi02}. In these works, $\phi$ and $g_n$ are functions of $n$ Ising variables, so that they  can take $2n$ different values. Below, we will replicate the $L_a$ so that they take $nN$ different values; $N=2$ is the Ising case.}. Indeed, any $L_a$ node can either propagate into $L_a^\dagger$ and become a leaf, with weight $g \xi O_a$, or propagate into $L_a^\dagger$ and become another branch, with weight $g \eta \sum_{b,c} M_{abc} L_b R_c$, including all possibilities. This gives \eqref{SP1}. Similarly, any $L_a^\dagger$ node is either the child of a root, with weight $g \zeta h \delta_{b1}$, or the child of a branch, with weight $g \eta \sum_{a',a,c} M_{a'ac} (L_a^\dagger+R_a^\dagger) R_c$, including all possibilities. This gives \eqref{SP2}. 

For specific grammars, \eqref{SP1}-\eqref{SP7} can be explicitly analyzed. { Indeed, after writing these equations in terms of real variables, these take the form of `context-free schema' as defined in section VII.6.1 in \cite{Flajolet09}. This class of equations, mainly defined by the fact that coefficients in the right-hand side of \eqref{SP1}-\eqref{SP7} are positive, generically have square root singularities at their radius of convergence \cite{Flajolet09}. Moreover, they can be solved by iteration. 

These equations are also related to cavity equations, or belief propagation equations, in the literature of disordered systems \cite{Mezard09}. For any probabilistic model living on a tree, the cavity equations are a closed, self-consistent set of equations that can be used to compute the partition function. The variables of these equations, known as messages, are probability measures on the symbols of the graph. Here the symbols are the hidden symbols, hence each message can be considered a variable with a colour index $a = 1 \ldots N$. 
 In general, there are two messages per site: one going into the interaction, which here is a branch, and one outgoing. For a WCFG, the interaction at a branch depends only on the values of the hidden symbols there; there is no intrinsic site-to-site disorder. Therefore for large trees, one can seek a solution to the cavity equations that is independent of site, only depending on whether sites are the left or right children of their parent. In this case, the cavity equations become \eqref{SP1}-\eqref{SP7}, up to some normalization factors. Therefore the variables $L$ and $R$, which above were introduced as dummy variables to generate the diagrammatic expansion, in fact have their proper interpretation as {\it messages}: $L$ is a downward message, while $L^\dagger$ is an upward message. In an iterative scheme to solve the cavity equations, these variables have additional time indices, hence the dynamical aspect of their name; \eqref{SP1}-\eqref{SP7} are the fixed-point equations.}


In what follows, our interest, however, is in extracting the behavior of typical grammars in the case when $N$ is large. For this we need to choose an ensemble. 
\section{Grammar ensembles}

\begin{figure}[t!]
\includegraphics[width=\textwidth]{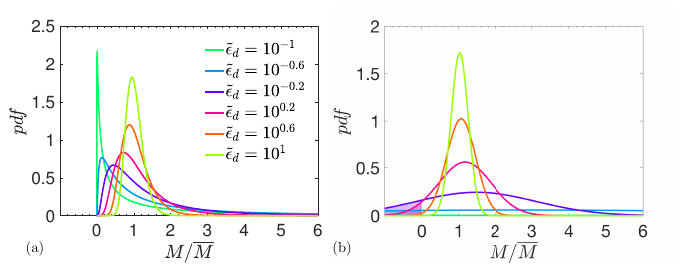}
\caption{Distribution of individual grammar weights in (a) Random Language model (RLM) and (b) Gaussian model for $\tilde\epsilon_d$ as indicated. The regime of unphysical negative weights is shaded in the Gaussian model. }\label{figens}
\end{figure}

We will consider two models. In \cite{DeGiuli19} it was argued that a generic model will have lognormally distributed weights, viz., 
\eq{ \label{PPG}
\PP_G(M,O) & \equiv Z_G^{-1} \; J \; e^{-\epsilon_d s_d}  e^{- \epsilon_s s_s } 
}
where the deep and surface sparsities $s_d$ and $s_s$ are defined by 
\eq{ \label{s1}
s_d = \frac{1}{N^3} \sum_{a,b,c} \log^2 \left[\frac{M_{abc}}{\overline{M}}\right], \;\; s_s =  \frac{1}{NT} \sum_{a,B} \log^2\left[\frac{O_{aB}}{\overline{O}} \right]
}
and $J = e^{-\sum_{a,b,c} \log M_{abc} - \sum_{a,B} \log O_{aB}}$. Here $\overline{M}=1/N^2$ and $\overline{O}=1/T$. A plot of the weights for a range of $\tilde\epsilon_d=\epsilon_d/N^3$ is shown in Fig.\ref{figens}a. It is straightforward to show that $\epsilon_d$ and $\epsilon_s$ satisfy
\eq{ \label{s2}
\overline{s_d} = (2\tilde \epsilon_d)^{-1}, \qquad \overline{s_s} = (2\tilde \epsilon_s)^{-1}.
}
where $\overline{\;\cdot\;}$ denotes a grammar average and $\tilde \epsilon_s = \epsilon_s/(NT)$. A small `deep temperature' $\epsilon_d$ corresponds to a large deep sparsity. The model \eqref{PPG} was called in \cite{DeGiuli19}  the Random Language Model (RLM).

It was shown in \cite{DeGiuli19} that the RLM shows two phases, depending on the value of $\tilde \epsilon_d$, plus logarithmic corrections. More precisely, Shannon entropies appear to collapse with respect to $\tilde \epsilon_d \log^k N$, where $k=1$ or $k=2$ depending on the quantity considered. For $\tilde \epsilon_d \log^k N \gtrsim 1$, Shannon entropies are independent of the deep temperature $\epsilon_d$, and take maximal values, indicating that the grammar does not carry information: despite strictly following the rules of a WCFG, sentences are indistinguishable from random noise. For smaller $\epsilon_d$, entropies drop, and the grammar carries nontrivial information. It is our goal to extract this transition from \eqref{W2}.

It will turn out to be much simpler to consider an alternative model, where the weights $M_{abc}$ and $O_{aB}$ are Gaussian, rather than lognormal; matching the mean and variance to those of the RLM we can again use the quantities $\overline{M}$ and $\tilde\epsilon_d$, and similarly for $O_{aB}$. The distribution is plotted in Fig.\ref{figens}b. This model has the unphysical feature that weights have a negative tail; naively, we could imagine that this would be unimportant, since the largest weights are most important, but we will have to revise this statement later. We call this the Gaussian model (GM).

We wish to compute
\eq{
\overline{\log \ZZ(\GG)} = \overline{\left. \frac{\p \ZZ(\GG)^n}{\p n} \right|_{n=0}} = \left. \frac{\p}{\p n} \right|_{n=0}  \overline{\ZZ(\GG)^n},
}
where we used the replica method \cite{Mezard87}. The fields $\zeta,\xi,\eta, L$, and $R$ are all replicated, adding an index $i=1 \ldots n$. To compute $\overline{\FF(\GG)^n}$ we need the grammar average
\eq{
A(\xi,\eta,L,R) & = \overline{ e^{\xi \sum_{i,a} O_a H_a^i + \eta \sum_{i,a,b,c} M_{abc} H_a^i L^i_b R^i_c} } \notag \\
& = \prod_{a,B} \overline{ e^{\xi  O_{aB} \sum_{i} H_a^i} } \prod_{a,b,c} \overline{ e^{\eta M_{abc} \sum_{i} H_a^i L^i_b R^i_c} }
}
Write $x_{abc} = \sum_i \eta_i H_a^i L^i_b R^i_c$. In the RLM the grammar averages over $M$ are of the form ($m=\log M_{abc}/\overline{M}$)
\eq{
\sqrt{\tilde \epsilon_d/\pi} \int dm \; e^{-\tilde\epsilon_d m^2} e^{e^m \overline{M} x} & = \sqrt{\tilde \epsilon_d/\pi} \int dm \; e^{-\tilde\epsilon_d m^2} \sum_{q \geq 0} \frac{\overline{M}^q x^q}{q!} e^{q m} \\
& = \sum_{q \geq 0} \frac{\overline{M}^q x^q}{q!} e^{q^2/(4 \tilde\epsilon_d)}
}
A term of order $q$ corresponds to the rule appearing $q$ times. We are interested in a transition due to patterns of repeated rule application between sentences, rather than inside them (this would correspond to a transition deeper in the ordered phase). Therefore {\it a priori} we expect that we only need {\it connected} terms up to a small finite order $q_*$. Note that at order $q_*$, terms involving $q_*$ different replicas will be present. Resumming gives
\eq{
\sum_{q \geq 0} \frac{\overline{M}^q x^q}{q!} e^{q^2/(4 \tilde\epsilon_d)} = \exp \left( \sum_{q \geq 1} a_q \overline{M}^q x^q \right) 
}
with $a_1 = e^{1/(4 \tilde\epsilon_d)}, a_2 = \half ( e^{4/(4 \tilde\epsilon_d)} - e^{2/(4 \tilde\epsilon_d)} ),$ etc. From this divergent sum we only need to retain terms up to order $q=n(\ell-m)$, since the integration over $\eta$ will retain only $\ell-m$ vertices for each replica.

For practical reasons exact calculations are limited to $q_*=2$. In this case, we consider all derivations in which a rule can appear at most twice in one derivation tree. Note that rules can still appear arbitrarily many times in the set of $n$ replicas and $m$ sentences. Keeping terms to $q_*=2$ is equivalent to letting the $M$ be drawn from a Gaussian distribution. For appropriate choice of mean and variance, we can thus fix the GM to be equal to the RLM to this order. In the remainder of this work, we will first find the exact solution of the GM, and then discuss its extension to the RLM. 

\subsection{Gaussian model} 


Applying the same arguments to the integral over $O$, we have for the GM
\eq{
A(\xi,\eta,L,R) & = \prod_{a,B} e^{ b_1 \overline{O} \sum_i \xi_i H_a^i + b_2 \overline{O}^2  (\sum_{i} \xi_i H_a^i )^2} \prod_{a,b,c} e^{a_1 \overline{M} \sum_{i} \eta_i H_a^i L^i_b R^i_c + a_2 \overline{M}^2 (\sum_{i} \eta_i H_a^i L^i_b R^i_c )^2} \notag \\
& = e^{ b_1 N T \overline{O} \sum_i \xi_i H_*^i + b_2 T \overline{O}^2 \sum_{i,j} \xi_i \xi_j Q_H^{ij} + a_1 N^3 \overline{M}  \sum_{i} \eta_i H_*^i L_*^i R_*^i + a_2 \overline{M}^2  \sum_{i,j} \eta_i \eta_j Q_H^{ij} Q_L^{ij} Q_R^{ij}}
}
where we introduced `magnetization' vectors and overlap matrices
\eq{
H_*^i = \frac{1}{N}\sum_a H_a^i, \quad L_*^i = \frac{1}{N}\sum_a L_a^i, \quad R_*^i = \frac{1}{N}\sum_a R_a^i, \\
Q_H^{ij} = \sum_a H_a^i H_a^j, \quad Q_L^{ij} = \sum_a L_a^i L_a^j, \quad Q_R^{ij} = \sum_a R_a^i R_a^j,
}
and $b_1 = e^{1/(4 \tilde\epsilon_s)}, b_2 = \half ( e^{4/(4 \tilde\epsilon_s)} - e^{2/(4 \tilde\epsilon_s)} )$. (Recall that $H_a = L_a^\dagger + R_a^\dagger$.). Assembling the above results we find that for the GM,
\eq{
\overline{\ZZ(\GG)^n} & = \prod_i \left[m!\oint' \frac{d\zeta_i}{\zeta_i^{1+m}} \oint' \frac{d\xi_i}{\xi_i^{1+\ell}} \oint' \frac{d\eta_i}{\eta_i^{1+\ell-m}} \int DL^i \int DR^i \; e^{-\frac{1}{g} \sum_a \left[ L^i_a L^i_a{}^\dagger + R^i_a R^i_a{}^\dagger \right] } \right] \notag\\
&\qquad e^{ \sum_{i} \zeta_i h (L_1^i + R_1^i)} A(\xi,\eta,L,R) \notag
} 
This is now in the form amenable to standard treatment by replicas: the overlap matrices can be introduced as new parameters, and the original variables can be integrated out. We notice that the colour indices play the role usually played by spatial indices in spin glasses \cite{Mezard87}. The surprising result is that the model can be exactly integrated, without even making an {\it ansatz} on the replica structure, and without taking the large $\ell$ limit. This integrability can be traced to the fact that the overlap matrices depend only on the real and imaginary parts of $L$ in the canonical way, i.e. through $L_a^i = Re[L_a^i] + i Im[L_a^i]$. This gives rise to a symplectic structure that simplifies the integration over these variables. The derivation is sketched in Appendix A. The final result is
\eq{ 
\overline{\log \ZZ} & = m \log h +(2\ell-m) \log g + S_{\ell,m} + \ell \log (T \Obar b_1) + \lbar \log (N^2 \Mbar a_1) \label{F1} \\
& + \ell \;f(x_s) + \lbar f(x_d) \notag
}
with $\lbar = \ell-m$,
\eq{ \label{fx}
f(x) = \int \frac{dt}{\sqrt{2\pi}} \; e^{-\frac{1}{2}t^2} \; \log\big[1+ xt\big],
}
and 
\eq{
S_{\ell,m} = \log \lbar! - \log \ell! + \log (1+\ell+\lbar)! - \log (1+2\lbar)!
}
Here $x_d=N^{-3/2} \sqrt{e^{1/(2\tilde \epsilon_d)}-1}$ and $x_s = (NT)^{-1/2} \sqrt{e^{1/(2\tilde \epsilon_s)}-1}$. 

\begin{figure}[t!]
\includegraphics[width=0.5\textwidth]{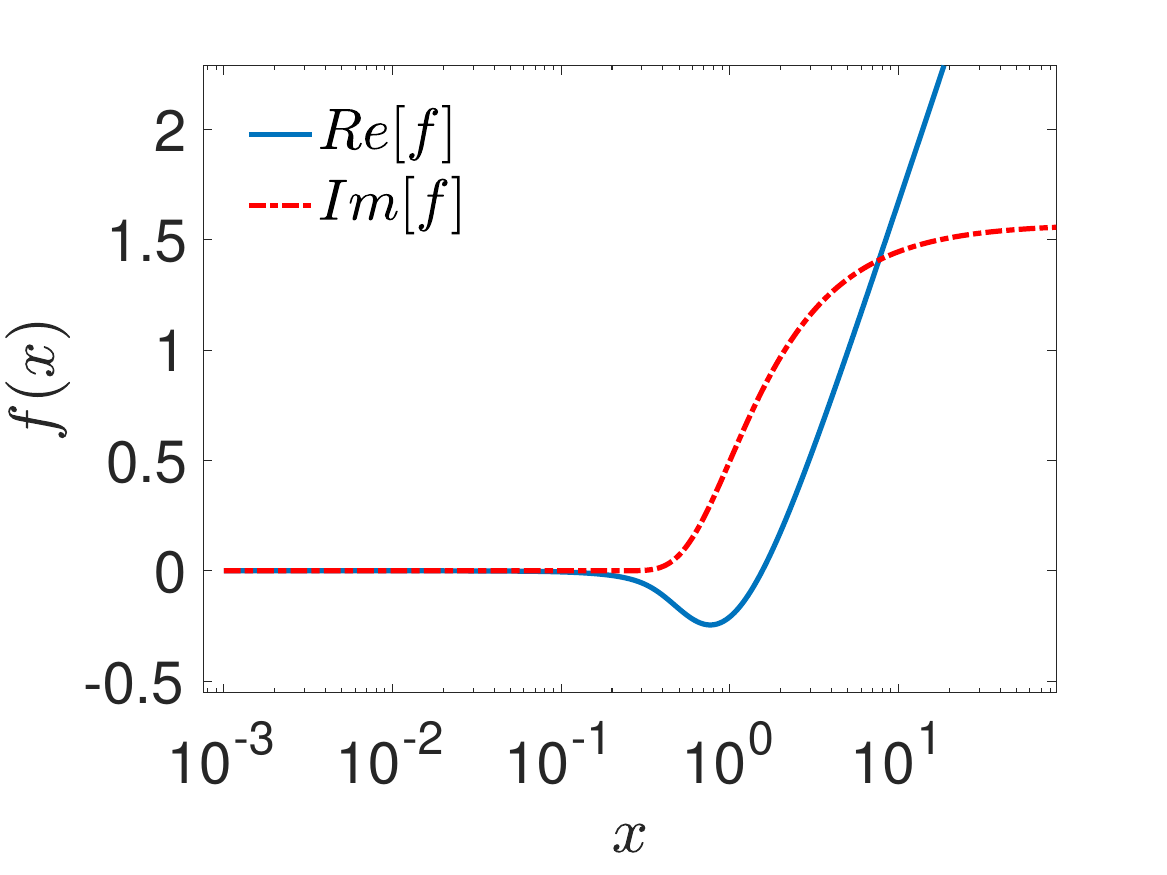}
\caption{Function $f(x)$ from \ref{fx}. }\label{figfx}
\end{figure}

The elements of $\overline{\log \ZZ}$ are as follows. { Those involving $h$ and $g$ give $\log \ZZ_{tree}$, the contribution to $\log \ZZ$ that weights configurations by the number of sentences and leaves; these are precisely those required from \ref{diagweight}. $S_{\ell,m}$ is entropic, and apparently counts the number of ways to partition a total string of length $\ell$ into $m$ sentences. } 
 Terms with $b_j$ and $a_j$ are energetic, since they depend on the grammar weight distribution. { Those involving $b_1$ and $a_1$ capture the change in the mean occupancy as temperature is varied. The function $f(x)$ captures the non-trivial effects of correlation between different symbols. This function, which can written in terms of hypergeometric functions, is plotted in Fig.\ref{figfx}. } It develops an imaginary part for $x \sim \OO(1)$, indicating that the unphysical negative probability states are becoming important. For large $N$ the condition $x_d \lesssim \OO(1)$ is equivalent to
\eq{ \label{ineq1}
\tilde \epsilon_d \log N \gtrsim 1/6,
}
and similarly the condition $x_s \lesssim \OO(1)$ is equivalent to $\tilde \epsilon_s \log NT \gtrsim 1/2$. These inequalities fix the regime in which the GM is physical. 

\subsection{RLM} We now return to the full model. As discussed above, we cannot obtain the exact solution; however, since a saddle-point method is justified for large $\ell$, we can consider different {\it ansatze} on the form of the solution. Two are natural: (i) the colour-symmetric {\it ansatz} $L_a^i = L^i, R_a^i = R^i$ $\forall a$, and the (ii) replica-symmetric {\it ansatz} $L_a^i = L_a, R_a^i = R_a$ $\forall i$. After some calculations similar to those for the GM, we eventually find for either (i) or (ii) the same form \eqref{F1}, except that $f \equiv 0$. Besides the {\it ansatze} on the form of $L$ and $R$, we assume that $\ell$ is large and that the replica limit $n \to 0$ can be taken perturbatively, i.e. keeping terms $\OO(n)$ in the action, as in the usual approach \cite{Mezard87} \footnote{Consistency with the GM suggests that we should be able to recover \eqref{F1} with $f\equiv 0$ for that model, without necessarily taking the limit $\tilde \epsilon_d \to \infty, \tilde \epsilon_s \to \infty$, in which case the $f$ terms are trivially unimportant. Indeed, if instead of taking $n \to 0$ in \eqref{A1}, we look for a saddle-point with large $\ell$, the saddle-point perturbative in $n$ gives exactly \eqref{F1} with $f\equiv 0$. This indicates that the function $f$ is nonperturbative in the replica limit.}. We now analyze \eqref{F1} with $f\equiv 0$, with the understanding that this holds in the replica-symmetric regime, whose range of validity is to be determined. 

It is convenient to separate $F=-\overline{\log \ZZ}$ into its entropic and energetic contributions. This can be done exactly because the RLM has a scaling symmetry when the bias $\beta$ is included. Indeed, it is not hard to show that the partition function satisfies the scaling property $\overline{\ZZ^n}(\beta,\tilde \epsilon_d, \overline{M}, \tilde \epsilon_s, \overline{O}) = \overline{\ZZ^n}(1,\tilde \epsilon_d/\beta^2, \overline{M}^\beta, \tilde \epsilon_s/\beta^2, \overline{O}^\beta)$ (in abuse of earlier notation). The $\beta-$dependent part of $F$ is
\eq{
F_\beta = - \beta \ell \log \Obar - \beta \lbar \log \Mbar - \frac{\beta^2 \ell}{4 \tilde\epsilon_s} - \frac{\beta^2 \lbar}{4 \tilde\epsilon_d}
}
so that at $\beta=1$, $E = -\ell \log \Obar - \lbar \log \Mbar - \frac{\ell}{2 \tilde\epsilon_s} - \frac{\lbar}{2 \tilde\epsilon_d}$ and the replica-symmetric entropy is
\eq{
S_{RS} = \lbar \log (gN^2/h) + \ell \log (gTh) - \frac{\ell}{4 \tilde\epsilon_s} - \frac{\lbar}{4 \tilde\epsilon_d} + S_{\ell,m}
}
The entropy cannot be negative; this gives necessary, though perhaps not sufficient, conditions on the regime where the replica-symmetric {\it ansatz} is applicable. For simplicity, consider the case $\ell \to \infty, m/\ell=\alpha \ll 1$ ($\ell/m$ is the typical length of a tree; we let it be large). Then one can determine that $S_{\ell,m} = \OO(\alpha \ell)$ and, for the simulated case where the emission probability is close to $p=1/2$, $g=h\approx 1/\sqrt{8}$. The condition $S_{RS}>0$ is approximately equivalent to
\eq{
\frac{1}{4 \tilde\epsilon_d} + \frac{1}{4 \tilde\epsilon_s} \lesssim \log(N^2 T/8)
}
Our main concern is the emergence of deep structure, which does not depend on what happens at the surface of the tree. In the limit $\tilde \epsilon_s \to \infty$, this becomes $\tilde \epsilon_d \log (N^2 T/8) \gtrsim 1/4$, very similar to the regime in which the GM is physical, \ref{ineq1}.

\subsection{Order parameter}
Finally, we return to the order parameter $Q_2$ that measures deep structure. Let us first give a heuristic derivation of its value in the RLM. We need to compute $q_0 = \overline{\langle \pi_{abc} \rangle^2}$ and $q_1 = \overline{\langle\pi_{abc}\rangle\langle\pi_{ab'c'}\rangle}$, where $\pi_{abc}$ is the occupancy of the rule $a\to bc$. Using \eqref{E} and \eqref{pi} in the diagrammatic representation, one can see that $\pi_{abc} = e^{h_{abc}} \eta M_{abc} H_a L_b R_c$.  The $\pi_{abc}$ satisfy the sum rule $\sum_{abc} \pi_{abc} = \lbar$ and so have a mean value $\overline{\pi}=\lbar/N^3$. The occupancies are positively correlated with the grammar weights, since rules with higher weights are sampled more frequently. A crude estimate is then $\langle \pi_{abc} \rangle/\overline{\pi} \sim M_{abc}/(\Mbar a_1)$ (the mean value of a weight is $\Mbar a_1$). This leads to the estimate
\eq{ 
Q_2 &\sim \frac{N^5(N^2-1)}{\lbar^2} \frac{\lbar^2}{N^6 \Mbar^2 a_1^2} \overline{M_{abc}^2 - M_{abc} M_{ab'c'}} \notag \\
& = \frac{N^2-1}{N} \big( e^{1/(2\tilde \epsilon_d)}-1 \big), \label{Q2}
} 
which indicates that order increases as $\epsilon_d$ is lowered simply because the weight variance increases. $Q_2$ can be computed more precisely using replicas. After a long computation, assuming the replica symmetric {\it ansatz} and taking large $\ell$, one eventually finds exactly the same result, \eqref{Q2}. Thus this simple expression is in fact the genuine replica symmetric result; it is plotted in Fig.\ref{figQ2}, where it is compared with numerical data \footnote{These data have been obtained by the same methods as described in \cite{DeGiuli19}. Here we have simulated many more samples ($\ell \sim 10^5$ compared to $\ell \sim 10^3$ in that work) in order to resolve the large $\epsilon_d$ part of the curve.}. In the large $\ell$ limit, it matches quantitatively, without fitting parameters, above $\tilde \epsilon_d \log N \gtrsim 1$. For smaller $\tilde \epsilon_d$, the data asymptote, as they must, while the replica-symmetric prediction diverges.
 
\section{Conclusion}

We showed that the partition function for weighted context-free grammars has a convenient diagrammatic representation. For individual grammars, the behavior of a large text $\ell \gg 1$ is governed by saddle-point equations, which resemble belief-propagation equations \cite{Mezard09}. 

We then considered two ensembles of grammars, which are equivalent in a large temperature limit. The Gaussian model (GM) was solved exactly, and shown to become unphysical for $\tilde \epsilon_d \log N \lesssim 1/6$. For the random language model (RLM), previously simulated in \cite{DeGiuli19}, the partition function was computed in the replica-symmetric {\it ansatz}; the entropy becomes negative at low temperature, again depending essentially on the quantity $\tilde \epsilon_d \log N$. Finally, the order parameter $Q_2$ was computed in the replica-symmetric ansatz. The prediction quantitatively agrees with simulations above $\tilde \epsilon_d \log N \gtrsim 1$. These results indicate that replica-symmetry must be broken in the nontrivial low-temperature phase. 

The RLM bears some similarity to a spin-glass on the Bethe lattice, a difficult problem that is still not fully understood \cite{Mezard01,Parisi17}. Indeed, both problems can be generated by a diagrammatic method \cite{Baillie95}, and in both problems one finds that overlaps of all orders are needed to compute the partition function \cite{De-Dominicis89,Mezard01,Parisi02,Parisi17}. However, for the spin-glass, one can perform an expansion around the mean-field limit \cite{De-Dominicis89,Mezard01,Parisi02}, which is the Sherrington-Kirkpatrick model solved by Parisi \cite{Mezard87}. Na\"ively, the analogue to the SK model would be the Gaussian model, which we solved above. However, we showed that this model does not break the replica symmetry. This is related to a gauge symmetry in the diagrammatic formulation. It is therefore an open question whether there is a more primitive model that captures the essence of random languages in the low-temperature phase, and remains solvable. 

Finally, we have focussed here on context-free grammars, for which derivations are trees. The next level up in the Chomsky hierarchy are context-sensitive grammars. A theorem of Kuroda \cite{Kuroda64} says that it is sufficient to add rules of the form $ab\to cd$ to those above to model all context-sensitive grammars. Clearly this will add a quartic vertex to our \eqref{I}, which is not in itself a difficulty. However, well-formed derivations must be represented by planar diagrams, so that the order of symbols is preserved in the derivation. Generating random planar graphs that are not trees requires matrices as integration variables; this strongly suggests that general grammars require the full machinery of complex matrix models. 

\ack
I am grateful to V. Kazakov and G. Semerjian for conversations at an early stage of this work, and to G. Parisi and W. Bialek for encouragement.
\vfill

\bibliographystyle{iopart-num}
\bibliography{../../language}

\providecommand{\newblock}{}
\begin{thebibliography}{10}
\expandafter\ifx\csname url\endcsname\relax
  \def\url#1{{\tt #1}}\fi
\expandafter\ifx\csname urlprefix\endcsname\relax\def\urlprefix{URL }\fi
\providecommand{\eprint}[2][]{\url{#2}}

\bibitem{DeGiuli19}
DeGiuli E 2019 {\em Phys. Rev. Lett.\/} {\bf 122} 128301

\bibitem{Loreto07}
Loreto V and Steels L 2007 {\em Nature Physics\/} {\bf 3} 758 

\bibitem{Loreto11}
Loreto V, Baronchelli A, Mukherjee A, Puglisi A and Tria F 2011 {\em Journal of
  Statistical Mechanics: Theory and Experiment\/} {\bf 2011} P04006 
  1742--5468

\bibitem{Burridge17}
Burridge J 2017 {\em Physical Review X\/} {\bf 7} 031008

\bibitem{Carnie13}
Carnie A 2013 {\em Syntax: A generative introduction\/} (John Wiley and Sons,
  Ltd.)

\bibitem{Chomsky02}
Chomsky N 2002 {\em Syntactic structures\/} (Berlin: Walter de Gruyter)

\bibitem{Hopcroft07}
Hopcroft J~E, Motwani R and Ullman J~D 2007 {\em Introduction to automata
  theory, languages, and computation\/} 3rd ed (Boston, Ma: Pearson)

\bibitem{Lin17}
Lin H~W and Tegmark M 2017 {\em Entropy\/} {\bf 19} 299

\bibitem{Piattelli-Palmarini15}
Piattelli-Palmarini M and Vitiello G 2015 {\em Biolinguistics\/} {\bf 9}
  96--115

\bibitem{Gallego17}
Gallego A~J and Orus R 2017 {\em arXiv preprint arXiv:1708.01525\/}

\bibitem{Pestun17}
Pestun V and Vlassopoulos Y 2017 {\em arXiv preprint arXiv:1710.10248\/}

\bibitem{Searls02}
Searls D~B 2002 {\em Nature\/} {\bf 420} 211

\bibitem{Di-Francesco95}
Di~Francesco P, Ginsparg P and Zinn-Justin J 1995 {\em Physics Reports\/} {\bf
  254} 1--133 

\bibitem{Kazakov86}
Kazakov V 1986 {\em Physics Letters A\/} {\bf 119} 140--144

\bibitem{Boulatov87}
Boulatov D and Kazakov V 1987 {\em Physics Letters B\/} {\bf 186} 379--384 
  0370--2693

\bibitem{Bachas94}
Bachas C, De~Calan C and Petropoulos P 1994 {\em Journal of Physics A:
  Mathematical and General\/} {\bf 27} 6121

\bibitem{Baillie95}
Baillie C, Janke W, Johnston D and Plech{\'a}{\v c} P 1995 {\em Nuclear Physics
  B\/} {\bf 450} 730--752 

\bibitem{Bessis80}
Bessis D, Itzykson C and Zuber J~B 1980 {\em Advances in Applied Mathematics\/}
  {\bf 1} 109--157 

\bibitem{tHooft74}
t'Hooft G 1974 {\em Nuclear Physics. B\/} {\bf 72} 461--473

\bibitem{Brezin78}
Brezin E, Itzykson C, Parisi G and Zuber J 1978 {\em Commun. math. Phys\/} {\bf
  59} 35--51

\bibitem{Le-Guillou12}
Le~Guillou J~C and Zinn-Justin J 2012 {\em Large-order behaviour of
  perturbation theory\/} vol~7 (Elsevier)

\bibitem{Di-Francesco06}
Di~Francesco P 2006 {\em 2D quantum gravity, matrix models and graph
  combinatorics\/} (Springer) pp 33--88

\bibitem{Eynard16}
Eynard B 2016 {\em Counting surfaces\/} (Springer)

\bibitem{Parisi02}
Parisi G and Tria F 2002 {\em The European Physical Journal B-Condensed Matter
  and Complex Systems\/} {\bf 30} 533--541

\bibitem{Flajolet09}
Flajolet P and Sedgewick R 2009 {\em Analytic combinatorics\/} (cambridge
  University press)

\bibitem{Mezard09}
M\'ezard M and Montanari A 2009 {\em Information, physics, and computation\/}
  (Oxford University Press)

\bibitem{Mezard87}
M\'ezard M, Parisi G and Virasoro M 1987 {\em Spin glass theory and beyond: An
  Introduction to the Replica Method and Its Applications\/} vol~9 (World
  Scientific Publishing Company)

\bibitem{Mezard01}
M{\'e}zard M and Parisi G 2001 {\em The European Physical Journal B-Condensed
  Matter and Complex Systems\/} {\bf 20} 217--233

\bibitem{Parisi17}
Parisi G 2017 {\em Journal of Statistical Physics\/} {\bf 167} 515--542

\bibitem{De-Dominicis89}
De~Dominicis C and Goldschmidt Y 1989 {\em Journal of Physics A: Mathematical
  and General\/} {\bf 22} L775 

\bibitem{Kuroda64}
Kuroda S~Y 1964 {\em Information and Control\/} {\bf 7} 207--223

\bibitem{Mackey03}
Mackey D~S and Mackey N 2003 On the determinant of symplectic matrices Tech.
  Rep. 422 Manchester Centre for Computational Mathematics

\end{thebibliography}

\appendix
\section{Solution of Gaussian model}

We introduce $Q_L^{ij} = \sum_a L_a^i L_a^j$ as a new variable with a corresponding momentum $\Lambda_L^{ij}$, and similarly for $Q_R^{ij} = \sum_a R_a^i R_a^j$ and $P^{ij} = \sum_a L_a^i R_a^j$, with conjugate momenta $\Lambda_R$ and $\Lambda_P$, respectively. Let us write $x_a^i = Re[L_a^i], y_a^i = Im[L_a^i]$. The variables $\{ x_a^i, y_a^i \}$ are Gaussian, with a coupling matrix diagonal in colour. For each $a$, the coupling matrix is a $2n \times 2n$ matrix acting on $(x_a^1,\ldots x_a^n,y_a^1,\ldots,y_a^n)$, 
\eq{
Z = \begin{bmatrix} \delb-g \hat{\Lambda}_L & -g i \hat{\Lambda}_L \\
-g i \hat{\Lambda}_L & \delb+g \hat{\Lambda}_L \end{bmatrix},
}
where $\delb$ is the $n\times n$ identity matrix. It is easily verified that $Z$ is complex symplectic: $J = Z^t \cdot J \cdot Z$, where 
\eq{
J = \begin{bmatrix} 0 & \delb \\ -\delb & 0 \end{bmatrix}
}  
This implies that $Z^{-1} = J^{-1} \cdot Z^t \cdot J$ and, less obviously, $|Z|=1$ \cite{Mackey03}. Hence after integrating out $\{ L_a^i \}$ there is no nontrivial entropic term from $\log |Z|$, nor does there appear the inverse of $\Lambda_L$, as would naively be expected. In fact, after integrating out $\{ L_a^i \}$ and $\{ R_a^i \}$ the action remains {\it linear} in $\Lambda_L$, $\Lambda_R$, and $\Lambda_P$. Hence these can be immediately integrated out and we find that 
\eq{
Q_L^{ij} = N L_*^i L_*^j, \quad Q_R^{ij} = N R_*^i R_*^j, \quad P^{ij} = N L_*^i R_*^j
}
We find
\eq{
\overline{\ZZ(\GG)^n} & = \prod_i \left[m!\oint' \frac{d\zeta_i}{\zeta_i^{1+m}} \oint' \frac{d\xi_i}{\xi_i^{1+\ell}} \oint' \frac{d\eta_i}{\eta_i^{1+\ell-m}} \int dL_*^i \int dR_*^i \; e^{-\frac{N}{g} \sum_a \left[ L^i_* L^i_*{}^\dagger + R^i_* R^i_*{}^\dagger \right] } \right] \notag\\
&\qquad e^{ \sum_{i} \zeta_i h (L_*^i + R_*^i)} e^{\tilde b_1 \sum_i \xi_i H_*^i + \tilde b_2 \left( \sum_{i} \xi_i H_*^i \right)^2 + \tilde a_1 \sum_i \eta_i H_*^i L_*^i R_*^i + \tilde a_2 \left(\sum_{i} \eta_i H_*^i L_*^i R_*^i \right)^2} \notag
} 
with $\tilde b_j = NT\Obar^j b_j, \tilde a_j = N^3 \Mbar^j a_j$, $j=1,2$. The quadratic terms can be linearized with a Hubbard-Stratonovich transformation, after which the integrals over $\zeta,\xi$, and $\eta$ are simple. The result is
\eq{ \label{A1}
\overline{\ZZ(\GG)^n} & = \frac{1}{2\pi}\int dp\;e^{-\frac{1}{2} p^2} \int dq\;e^{-\frac{1}{2} q^2}  \left[ \int dL \int dR \; K(L,R,p,q)\right]^n 
}
with
\eq{
K(L,R,p,q) & = \frac{h^m (\pi g/N)^{-2}}{\ell!(\ell-m)!} \big( \tilde b_1 + p \sqrt{2 \tilde b_2} \big)^\ell \left( \tilde a_1 + q \sqrt{2\tilde a_2} \right)^{\ell-m} \notag \\
& \qquad e^{-\frac{N}{g} \left[ L L^\dagger + R R^\dagger\right]} \left( L^\dagger + R^\dagger \right)^{2\ell-m} (LR)^{\ell-m} 
}
The calculation is thus reduced to an effective single-colour problem. We have
\eq{
& \frac{h^m (\pi g/N)^{-2}}{\ell!(\ell-m)!} \int dL \int dR \; e^{-\frac{N}{g} \left[ L L^\dagger + R R^\dagger\right]} \left( L^\dagger + R^\dagger \right)^{2\ell-m} (LR)^{\ell-m} \notag \\
& \qquad = \frac{h^m}{\ell!(\ell-m)!} (g/N)^{2\ell-m} \sum_{k=0}^m {m\choose k} (\ell-m+k)! (\ell-k)! \notag \\
& \qquad = \frac{h^m}{\ell!} (g/N)^{2\ell-m} (\ell-m)! \frac{(1+2\ell-m)!}{(1+2\ell-2m)!}
}
so that the final result is
\eq{
\overline{\log \ZZ} & = m \log h +(2\ell-m) \log (g/N) + S_{\ell,m} + \ell \log \tilde b_1 + (\ell-m) \log \tilde a_1 \notag \\
& + \ell \;f\left(\sqrt{2\tilde b_2/\tilde b_1^2}\right) + (\ell-m) f\left(\sqrt{2\tilde a_2/\tilde a_1^2}\right) 
}
with
\eq{
f(x) = \int \frac{dt}{\sqrt{2\pi}} \; e^{-\frac{1}{2}t^2} \; \log\big[1+ xt\big]
}
and $S_{\ell,m} = \log (\ell-m)! - \log \ell! + \log (1+2\ell-m)! - \log (1+2\ell-2m)!$. Letting $x_d=\sqrt{2\tilde a_2/\tilde a_1^2} = N^{-3/2} \sqrt{e^{1/(2\tilde \epsilon_d)}-1}$ and $x_s = (NT)^{-1/2} \sqrt{e^{1/(2\tilde \epsilon_s)}-1}$, we have the result shown in the main text.

\section{Corrigendum to J. Phys A: Math. Theor. 52 (2019) 504001}

{\it The above preprint proposed a diagrammatic formulation of the Random Language Model; explained why the model is dominated by saddle-points; and sought the solution to the disorder-averaged model by comparison to a simpler, solvable model. We discuss a hidden assumption of the latter analysis above that was neither explained nor motivated: the analytical solution to the Gaussian model, and its extension to the Random Language Model, are predicated on a ``downwards'' approximation that neglects information flow from the leaves to the root of derivation trees. }

\newcommand{\Ld}{{L^{\dagger}}}
\newcommand{\Rd}{{R^{\dagger}}}
\newcommand{\Qd}{{Q^{\dagger}}}
\newcommand{\Pd}{{P^{\dagger}}}
\newcommand{\Lamd}{{\Lambda^{\dagger}}}

The diagrammatic formulation of the Random Language Model (RLM) \cite{DeGiuli19} uses complex-valued fields $L_a^i$ and $R_a^i$ where $a = 1 \ldots N$ is a colour index and $i=1 \ldots n$ is a replica index. Above we proposed a replica-symmetric solution to this model, using first a Gaussian model (GM) where the analysis is simplified. The solution to the GM was sketched in the Appendix. Here we point out a hidden assumption of the analysis above. 

Write $L_a^j = x_a^j + i y_a^j$ and $R_a^j = z_a^j + i w_a^j$. After taking the disorder average, the GM depends nonlinearly on several order parameters
\eq{
Q_L^{ij} & = \sum_a L_a^i L_a^j  \\ 
Q_R^{ij} & = \sum_a R_a^i R_a^j \\ 
Q_H^{ij} & = \sum_a (\Ld_a^i+\Rd_a^i) (\Ld_a^j+\Rd_a^j) \\
L_*^i & = \frac{1}{N} \sum_a L_a^i \\
R_*^i & = \frac{1}{N} \sum_a R_a^i \\
H_*^i & = \frac{1}{N} \sum_a (\Ld_a^i+\Rd_a^i),
}
The strategy for integrating the model is to introduce these order parameters as new integration variables. In so doing the original variables $L$ and $R$ are decoupled and can be integrated out. Since these are all complex-valued parameters, we need to treat their real and imaginary parts separately. For example, we can introduce $Q_L$ as $Q_{L,R} + i Q_{L,I}$ with 
\eq{
1 \propto \int D\Lambda_{L,R} \int DQ_{L,R} \int D\Lambda_{L,I} \int DQ_{L,I} e^{-i \sum_{i,j} \Lambda_{L,R}^{ij} [Q_{L,R}^{ij}-\sum_a [x_a^i x_a^j - y_a^i y_a^j] ]} e^{-i \sum_{i,j} \Lambda_{L,I}^{ij} [Q_{L,I}^{ij}-\sum_a [y_a^i x_a^j + x_a^i y_a^j]] } 
}
$Q_R$, $L_*$, and $R_*$ can be introduced in a similar manner. We can write $Q_H = \Qd_L + \Qd_R + \Pd + \Pd^T$ with 
\eq{
P^{ij} & = \sum_a L_a^i R_a^j \notag \\
& = \sum_a (x_a^i + i y_a^i) (z_a^j + i w_a^j) 
}
and then introduce $P$ as a new variable, again in real and imaginary parts. After introducing these quantities the partition function takes the form
\eq{
\overline{Z^n} = \int D[...] e^{-S_0(Q_L,Q_R,P,L_*,R_*,H_*)} e^{-S_1(\text{all})} e^{-S_2(\text{all}\;\Lambda, \text{all } \lambda)}
}
where $D[...]$ includes all integrations, including those over $\zeta$, $\xi$, and $\eta$ (see above).
The action $S_0$ contains the nontrivial RLM-specific dependence on the overlaps and `magnetization' vectors. The action $S_1$ contains the inner products of the overlap matrices and magnetization vectors with their corresponding Lagrange multipliers:
\eq{
S_1 = i \Lambda_{L,R} : Q_{L,R} + i \Lambda_{L,I} : Q_{L,I} + \ldots
}
Finally the action $S_2$ includes the integrals over the $\{x_a^i,y_a^i,z_a^i,w_a^i\}$, which can be done explicitly. It is a function of all the Lagrange multipliers. The remaining integrals in $\overline{Z}$ are determined by saddle-point equations. 

To simplify the problem, one can make an ansatz to reduce the problem complexity (for example that overlap matrices are of hierarchical (Parisi) type). Alternatively, here we have real and imaginary parts of variables and can make an ansatz on this structure. 

To motivate approximations, we recall from above that the diagrammatic formulation is related to the cavity method of disordered systems \cite{Mezard87}. $L$ has the interpretation of a downward message (from the root to the leaves), while $\Ld$ has the interpretation of an upward message (from the leaves to the root). Therefore variables in the action can be classified by whether they propagate information upwards (made from $\Ld$ and/or $\Rd$) or downwards (made from $L$ and/or $R$). Because of the model structure, it is natural that downwards messages are more important for observables that depend on the bulk variables (as opposed to the surface, which is formed by the leaves of the tree). Indeed the probability of any bulk observable can be written as a nontrivial function of the $M$ grammar only, multiplied by the normalization constant $1/Z$. The surface properties only affect probabilities through the normalization $Z$. 

Suppose, for the sake of argument, that we neglect the upwards information transfer. This is equivalent to neglecting the $\Qd_L$, $\Qd_R$, and $\Pd$ dependence in $S_0$, but leaving the $Q_L$, $Q_R$, and $P$ dependence. Consider then the saddle-point equation
\eqs{
0 = \frac{\p S}{\p \Qd_{L}^{ij}} & = \frac{\p S}{\p Q_{L,R}^{ij}} \frac{\p Q_{L,R}^{ij}}{\p \Qd_{L}^{ij} } + \frac{\p S}{\p Q_{L,I}^{ij}} \frac{\p Q_{L,I}^{ij}}{\p \Qd_{L}^{ij} } \\
 & = \frac{\p S}{\p Q_{L,R}^{ij}} \half + \frac{\p S}{\p Q_{L,I}^{ij}} (-\ffrac{1}{2i}) \\
 & = \half \left[ \frac{\p S}{\p Q_{L,R}^{ij}} + i \frac{\p S}{\p Q_{L,I}^{ij}} \right] \\
 & = \half \left[ \frac{\p S_0}{\p Q_{L,R}^{ij}} + i \Lambda_{L,R}^{ij}  + i\frac{\p S_0}{\p Q_{L,I}^{ij}} + i^2 \Lambda_{L,I}^{ij}\right].
}
Independence of $S_0$ with respect to $\Qd$ implies $\frac{\p S_0}{\p Q_{L,R}^{ij}} + i\frac{\p S_0}{\p Q_{L,I}^{ij}} = 0$ so that
\eq{
 \Lambda_{L,R}^{ij} + i \Lambda_{L,I}^{ij} = 0,
}
or $\Lambda_L=0$, schematically. Note that $\Lambda_L^\dagger = 2 \Lambda_{L,R} \neq 0$.

Applying this condition $\Lambda_{L}^\dagger$ appears in the action as
\eq{
& i \Lambda_{L,R} : [Q_{L,R} - Q_{L,R}[L]]+ i \Lambda_{L,I} : [Q_{L,I} - Q_{L,I}[L]] \notag \\
& i \Lambda_{L,R} : \left[ [Q_{L,R} - Q_{L,R}[L]] + i [Q_{L,I} - Q_{L,I}[L] ] \right] \notag \\
& \half i \Lambda_{L}^\dagger : \left[ Q_{L} - Q_{L}[L]  \right]
}
which is the form that was used above.

As shown above, the resulting theory correctly predicts the behaviour of the order parameter $Q_2$ in the high-temperature regime of the model. This suggests {\it a posteriori} that the ``downwards" approximation is justified in this regime. We expect that it breaks down in the nontrivial low-temperature regime.

This ``downwards" approximation leads to a symplectic propagator. To see this, consider the inverse propagator for the $L$ variables, i.e.
\eq{
M^{ij}_{ab} & = \frac{2}{g} \delta_{ab} \begin{bmatrix}  \delta^{ij} & 0\\ 0& \delta^{ij} \end{bmatrix} -  2i \delta_{ab} \begin{bmatrix}  \Lambda_{L,R}^{ij} & \Lambda_{L,I}^{ij} \\ \Lambda_{L,I}^{ij} & - \Lambda_{L,R}^{ij}\end{bmatrix} \notag \\
& \to \frac{2}{g} \delta_{ab} \delta^{ij} \begin{bmatrix} 1  & 0\\ 0 & 1 \end{bmatrix} -  2i \delta_{ab} \Lambda_{L,R}^{ij}  \begin{bmatrix} 1  &  i \\ i & -1\end{bmatrix} & \quad \text{downwards approx.}
}
which now has a partially factorized colour, replica, and real/imaginary dependence. 
Define the replica part $\tilde M^{ij}$ by $M^{ij}_{ab} = (2/g) \delta_{ab} \tilde M^{ij}$, i.e.
\eq{
\tilde M = \begin{bmatrix} \delb - i g' \Lamd_L  &  g' \Lamd_L\\ g' \Lamd_L & \delb + i g' \Lamd_L \end{bmatrix},
}
where $g'=g/2$ and $\delb$ is the $n\times n$ identity matrix. Let 
\eq{
J \equiv \begin{bmatrix}  0 & \delb \\ -\delb & 0 \end{bmatrix} 
}
and note that 
\eq{
\tilde M^T \cdot J \cdot \tilde M & = \begin{bmatrix} \delb - i g' \Lamd_L  &  g' \Lamd_L\\ g' \Lamd_L & \delb + i g' \Lamd_L \end{bmatrix} \cdot \begin{bmatrix} g' \Lamd_L  &  \delb + i g' \Lamd_L \\ -\delb + i g' \Lamd_L & -g' \Lamd_L \end{bmatrix} \notag \\
& = \begin{bmatrix} g'\Lamd_L - i g'^2 \Lamd_L^2 - g' \Lamd_L + i g'^2 \Lamd_L^2 &  \delb - i^2 g'^2 \Lamd_L^2 - g'^2 \Lamd_L^2\\ g'^2 \Lamd_L - \delb + i^2 g'^2 \Lamd_L^2 & g' \Lamd_L + i g'^2 \Lamd_L^2 - g' \Lamd_L - i g'^2 \Lamd_L^2 \end{bmatrix} \notag \\
& = J
}
which means that $\tilde M$ is complex symplectic. This implies that $\tilde M^{-1} = J^{-1} \cdot \tilde M^T \cdot J$ and also $\det \tilde M=1$. These relations simplify the analysis. 

We note in passing that the ``upwards'' approximation that neglects downwards information flow would also lead to a (different) symplectic propagator. 

The analysis of the GM (and by extension the RLM) without making this ``downwards'' approximation will be reported elsewhere.

\end{document}